\documentclass[fleqn,10pt]{wlscirep}
\usepackage[utf8]{inputenc}
\usepackage[T1]{fontenc}

\usepackage{soul}

\title{One-particle engine with a porous piston}

\author[1,*]{Carlos E. Álvarez}
\author[2,+]{Manuel Camargo}
\author[3,+]{Gabriel Téllez}
\affil[1]{Escuela de Ingenier\'\i a, Ciencia y Tecnolog\'\i a, Universidad del Rosario, Bogotá, Colombia}
\affil[2]{FIMEB \& CICBA, Universidad Antonio Nariño - Campus Farallones, Cali, Colombia}
\affil[3]{Departamento de Física, Universidad de los Andes, Bogotá, Colombia}

\affil[*]{carlosedu.alvarez@urosario.edu.co}
\affil[+]{these authors contributed equally to this work}

\begin{abstract}
 We propose a variation of the classical Szilard engine that uses a porous piston. Such an engine requires neither information about the position of the particle, nor the removal and subsequent insertion of the piston when resetting the engine to continue doing work by lifting a mass against a gravitational field. Though the engine operates in contact with a single thermal reservoir, the reset mechanism acts as a second reservoir, dissipating energy when a mass that has been lifted by the engine is removed to initiate a new operation cycle. 
\end{abstract}
\begin{document}

\flushbottom
\maketitle
%
%
\thispagestyle{empty}


\section*{Introduction} 

In 1929 Leo Szilard proposed a version of Maxwell's demon that appeared to violate the Kelvin-Planck statement of the second law by lifting a mass against a gravitational field while in contact with a single heat reservoir~\cite{Szilard1929}. The demon consists of an engine working with a single particle gas, and since then several models for single particle engines and processes have been studied (see for example~\cite{Hoppenau2013,Proesmans2015,Alvarez2019}). Though some variants of this demon, known as Szilard's engine, have been proposed~\cite{Leff2003,Bennett1987,Marathe2010,Kim2011,Boyd2016,Bhat2019}, the basic design is composed of the following parts: A particle inside a container along with a removable piston, a heat bath at temperature $T$ in contact with the container
walls, a hanging mass which is attached to the piston in such a way that work is performed on it as the piston moves, and an external agent in charge of resetting the system once the piston has reached one of the ends of the container.\\

During a cycle, the agent inserts the piston in the middle of the container and measures the particle's position with respect to the piston (left or right). Next, it attaches the mass to the piston from the same side as the particle, such that when the piston is released the particle pushes it to the opposite end of the container, simultaneously lifting the mass in the process. Finally, the agent removes the piston from the end of the container, taking away with it the potential energy gained by the mass in the current cycle and preparing the system for the next one.\\

However, the agent generates entropy with its actions. The information obtained from the measurement is stored in a memory and it is used when the decision to hang the mass, either from the left or right, has to be made. Landauer found that in order to reset the memory an amount of entropy equal to the entropy reduction produced by the engine has to be generated \cite{Landauer1961}. To this minimal entropy one can add other terms, like the amount of entropy generated when resetting the piston, which has also been calculated \cite{Kish2012}. This connection between physics and information theory is now the subject of the field of thermodynamics of information \cite{Parrondo2015}.\\

Currently, there is a growing interest in automating the process by which a small out-of-equilibrium system like Szilard's engine generates work, by replacing a demon that acquires and processes information by an autonomous agent which is coupled to the system \cite{Kutvonen2016,Sanchez2019,Ciliberto2020}. The second law of thermodynamics still holds in such schemes, as the autonomous agent has to be coupled to a thermal reservoir at a temperature which is lower than the lowest temperature in the system \cite{Freitas2021}.\\

In this paper we propose a variation of Szilard's model engine that does work without the need for an agent to measure the position of the particle, to decide based on this information from which side of the container to hang the mass, or to reallocate the piston in the reset step of the process. This system is not driven by changing a set of variables following some defined protocol and, according to our analysis, it still respects the Kelvin-Planck statement of the second law, as even if the engine is technically in contact with a single thermal reservoir the process dissipates energy through its reset mechanism.\\

\section{Model}{\label{modSec}}

\begin{figure*} 
  \centering
  \includegraphics[scale=0.36]{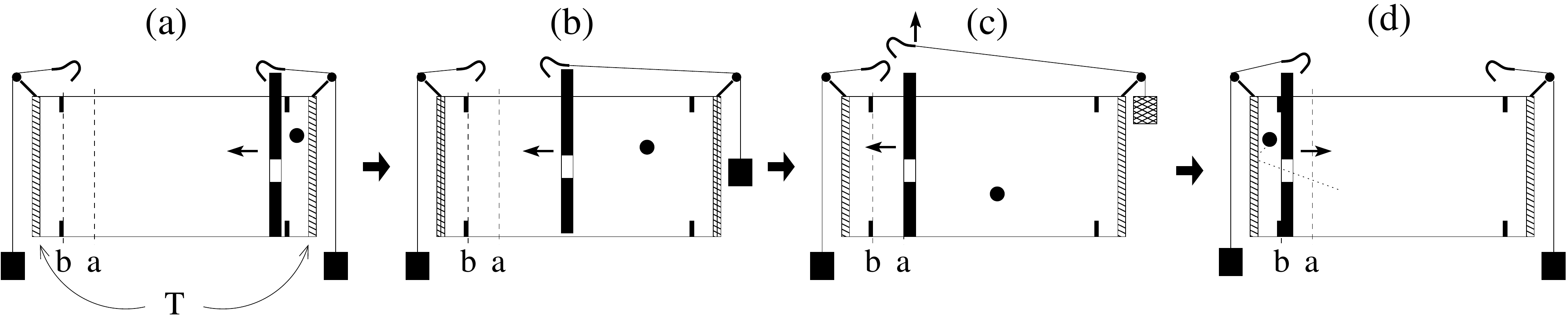}
  \caption{Schematic of the engine with porous piston. (a) The piston is initially connected to the right load mass and the particle bounces between the right thermal wall and the piston. (b) The particle pushes the piston to the left. (c) Once the piston gets to position $a$, the high energy load mass hanging from its right is removed and a new low energy one is placed on the right end of the container. (d) As the piston gets to point $b$ it couples with the left load mass. Once the particle crosses to the left side through the pore it will push the piston to the right and the cycle repeats.}
  \label{engine1}
\end{figure*}

We propose a model composed by a single particle of mass $m$ and a piston of mass $M$ inside a one dimensional container of length $L$ whose ends are in contact with a heat bath at temperature $T$, as schematically shown in Fig.~\ref{engine1}. The piston has pores in it such that when the particle is close enough, there is a probability $p$ for it to cross to the other side of the container instead of colliding with the piston. We refer to $p$ as the {\it porosity} of the piston. As with Szilard's model, a load mass $M_l$ can be hanged from the piston so that when it is pushed by the particle, the mass is lifted. If $p$ is small enough, most of the time the particle will be able to push the piston to the opposite side of the container before crossing through the pore and lift the load mass in the process. Once the particle crosses, it will push the piston in the opposite direction.\\

An external agent is charged with the task of removing the incoming high potential energy mass once the piston has reached one of the container's ends, while another agent
prepares a new mass with low potential energy on the opposite side of the container, which will be coupled to the piston at a later instant. 
One can use feedback by measuring the state of the system to pinpoint the moment at which a mass has been removed from the system and use this information to place a new mass as soon as possible. 
To model the coupling between the piston and the hanging mass we assume that the mass is connected through an ideal string to a hook that the agent places at one of the ends of the container. Such a hook can couple to the piston by spending a small fixed amount of the kinetic energy of the piston. As the particle pushes the piston, the piston collides elastically with the mass hook, pushing it to the other side of the container until it gets removed past a given point $a$ (see Fig.~\ref{engine1}). When reaching point $b$, the piston hook couples with the mass hook at that point, dissipating once again the same arbitrary amount of kinetic energy.\\

The inner workings of the engine presented here can be understood in the context of nonequilibrium thermodynamics of feedback control~\cite{Sagawa2012}. Let us consider the system composed of the particle and the piston. Let $x$ be phase-space point of the system composed of the particle position and momentum and the piston position and momentum. The Hamiltonian of the system is composed of the interaction between the piston and the particle, the fixed external interaction between the particle and the container walls and the potential interaction $U_{pl}(x,\lambda)$ between the piston and the load masses. As the system evolves,  this interaction potential switches between three different forms depending on the load masses, labelled by $\lambda\in\{L,R,U\}$, corresponding respectively to a load mass on the left side ($L$), load mass on the right side ($R$) or  the piston is unhooked from the masses ($U$). If $\lambda$ is kept fixed, i.e., the unhook and hooking mechanism is switched off, the system will attain thermal equilibrium with an equilibrium free energy $F(\lambda)$. Due to the symmetry of reflection between left and right sides $F(L)=F(R)$,  therefore the difference of free energy between two states where a load mass is hooked is $\Delta F=0$.

When the hooking and unhooking mechanism is operating, a given function $\lambda(t)$ can be though as a protocol controlling the dynamics of the system. But, contrarily to many small-scale systems studied in stochastic thermodynamics~\cite{Sekimoto2010}, the protocol in this case is neither fixed nor deterministically applied to the system on each realization. Rather, the inner mechanisms of the engine described previously are responsible for switching $\lambda$ at random times, which can be interpreted as a feedback mechanism as follows. 

Suppose that at each time step, a measure without error is performed on the system to obtain $x$, from which the position of the piston and its kinetic energy are obtained. Then the following feedback control is performed:
\begin{itemize}
    \item If the piston is neither in any of the positions $a$ or $b$ shown in Fig.~\ref{engine1}, then $\lambda$ is unchanged from its current value. Note that there are actually two $a$ positions and two $b$ positions, located symmetrically at each side of the container: $a_L$ and $b_L$ at the left, and $a_R$ and $b_R$ at the right.
    \item If the piston is in a position $a_L$  and $\lambda=R$ or the piston is at $a_R$ and $\lambda=L$, then $\lambda$ is changed to $U$ (unhook). 
    \item If the piston is in a position $b_L$ (or $b_R$) and has enough kinetic energy, and $\lambda=U$ then $\lambda$ is changed to $L$ (or $R$) corresponding to the side the piston is at.
\end{itemize}

Similarly to Szilard's engine, there is an arbitrary amount of work that the external agents must perform to remove the masses that have been lifted and to bring new masses from the surroundings to the position at which they can be connected to the piston. We consider such agents as part of the surroundings and concentrate instead on the intrinsic capacity of the engine to perform work {\it given} that the protocol described previously is carried out by the agents. For simplicity, the new masses are introduced to the system with zero kinetic energy, as if they were drawn from a thermal reservoir at zero temperature.\\

It is not always possible to define cycles for engines at the nanoscale level in the same way as it is done macroscopically.\cite{Strasberg2021} However, in our case we find it useful to define a {\it cycle} as the process that occurs between the coupling of a new load mass to the piston and the subsequent removal of an old one. Note that the duration of such a cycle ($t_c$) is variable.\\

\section{Simulations}\label{simulations}

Mesoscale and smaller systems are commonly subject to diffusive effects and hydrodynamic interactions, as they are in general not found in vacuum. Such effects are important at the moment of thinking about a possible experimental realization. However, in this first study we are neglecting them to concentrate on the basic operation of a simple model of the engine.\\

We perform event driven simulations \cite{Allen1990} for ensembles of 200 systems using reduced mass ($M$), distance ($l$) and time ($t$) units defined as
\begin{align}
  M^*&=\frac{M}{m},\\
  l^*&=\frac{l}{L},\\
  t^*&=\sqrt{\frac{k_BT}{mL^2}}t,
\end{align}
which, for the energy ($E$) and acceleration ($g$) gives
\begin{align}
  E^*&=\frac{E}{k_BT},\\
  g^*&=\frac{mL}{k_BT}g.
\end{align}
For ease of notation the asterisks will be dropped in the following.\\

To simulate the interaction of the particle with the thermal walls, every time it collides with one of the container's end walls, its recoil speed $v$ is randomly drawn from the distribution \cite{Proesmans2015}
\begin{align}
  \rho(v)=\beta m v e^{-\frac{1}{2}\beta m v^2},
\end{align}
where $\beta=(k_BT)^{-1}$, $k_B$ is the Boltzmann constant and $T$ is the absolute temperature of the thermal walls.\\

The collisions between the particle and the piston are elastic and occur with probability $1-p$ every time they meet each other, otherwise the particle crosses through the pore and both the piston and the particle continue with their trajectories without interacting. Once the piston reaches a return point, indicated by $b$ in Fig.~\ref{engine1}, a small fixed amount of energy is subtracted from its kinetic energy to overcome the energy barrier to couple the hook.
The model assumes that this energy is dissipated. If the piston does not have enough kinetic energy then an elastic collision occurs. Lastly, as shown in Fig.~\ref{engine1}, once a load mass is coupled to the piston, it is pushed by means of elastic collisions until it gets pass point $a$, where it is removed from the system. Between collisions the load mass moves under the influence of the external gravitational field.\\

The energy dissipated every time a coupling occurs, as well as the kinetic energy that gets out of the system along with the masses and their hooks, is accumulated in a variable $E_{dis}$, that accounts for the energy dissipated by the process. The change in the gravitational potential energy of the load masses as they are lifted by the engine is accumulated in the variable $W$. The work done by the engine (particle + piston) on the load masses is $-W$. The change in kinetic energy of the particle each time a collision with one of the thermal walls occurs is accumulated in the variable $Q$, that accounts for the heat exchanged with the bath. Finally, the kinetic energy change of the piston during the collisions with the mass hooks is accumulated in the variable $W^{(ph)}$ which represents the work done by the hooks on the piston.\\

We compute the efficiency of the engine as the ratio between the power delivered by the engine on the hanging mass $\dot{W}_t$ during a time $t$ and the heat exchanged per unit of time $\dot{Q}_t$ during the same lapse
\begin{align}
  \eta_t=-\frac{\dot{W}_t}{\dot{Q}_t}.
\end{align}
In addition to the total efficiency $\eta_t$, we obtain the efficiency $\eta^{(bp)}_t$ of the energy transfer between the heat bath and the piston
\begin{align}
  \eta^{(bp)}_t=-\frac{\dot{W}^{(ph)}_t}{\dot{Q}_t},
\end{align}
as well as
\begin{align}
  \eta^{(pl)}_t=\frac{\dot{W}_t}{\dot{W}^{(ph)}_t},
\end{align}
that measures the conversion efficiency of work done by the piston on the hooks of the load masses into work done against the external field.\\

The parameters explored by the simulations where: $k_BT=1$, $M=50$, $b=0.075$, $a=0.15$ and the gravitational field $g=2$, along with several values for the porosity $p\in\{0.036,0.2,0.4,0.6,0.96\}$ and the load mass $M_l\in\{0.1,0.5,1.0,2.0,3.0\}$.\\


\section{Results}

\begin{figure}[!b]
 \centering
 \includegraphics[scale=0.055]{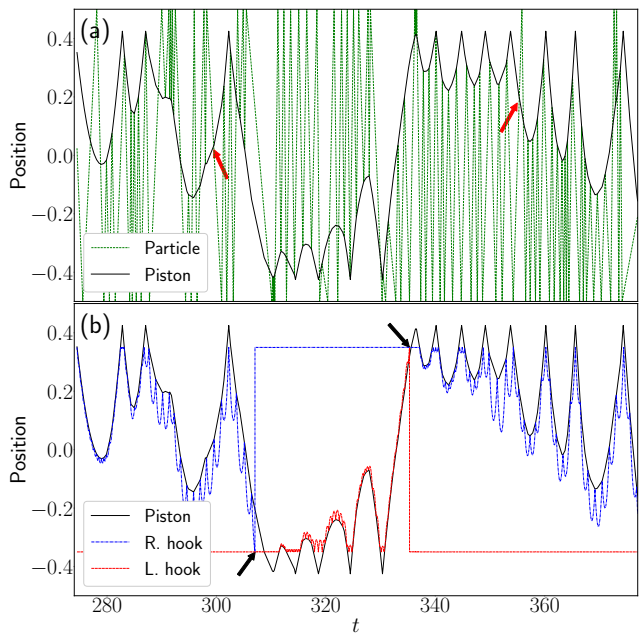}
 \caption{Example of the system dynamics. (a) Position along the container of the particle and the piston as a function of time. (b) Position of the piston, left and right hooks as a function of time, during a simulation of the engine where $p=0.2$ and $M_l=1.0$. The arrows in (a) point to some events where the particle crosses through the pore, while the arrows in (b) point to events where one of the hooks is pushed out of the system and reset.}
 \label{traject}
\end{figure}

As the engine is a system out of equilibrium featuring predominant fluctuations, we study the distribution of the efficiency \cite{Gingrich2014,Verley2014B} obtained from the simulations in order to estimate the macroscopic efficiency\cite{Verley2014,Proesmans2015,Proesmans2015B}
\begin{align}
  \bar{\eta}=-\frac{\langle\dot{W}\rangle}{\langle\dot{Q}\rangle},
\end{align}
where $\langle\cdot\rangle$ denotes the ensemble average in the thermodynamic limit ($t\rightarrow\infty$). To achieve this, we make use of the large deviation function\cite{Touchette2009,Touchette2011,Cover2006} defined as
\begin{align}
  J(\eta)=\lim_{t\rightarrow\infty}-\frac{1}{t}\ln P(\eta_t\in [\eta,\eta+d\eta]),
  \label{ldf}
\end{align}
where $P(\eta_t\in [\eta,\eta+d\eta])$ is the probability for $\eta_t$ to take a value between $\eta$ and $\eta+d\eta$. The large deviation function describes the asymptotic behavior ($t\rightarrow\infty$) of the efficiency fluctuations, and its minimum, which is located at the most probable value, indicates the macroscopic efficiency of the engine $\bar{\eta}$. To estimate $J(\eta)$, one can resort to the extrapolation procedure proposed by  Proesmans and van den Broeck\cite{Proesmans2015B}, which assumes a general functional form for $P_t(\eta)$ having three fitting parameters, that can be determined by using three evaluations of $P_t(\eta)$ at finite times.\\

An example of the dynamics of the different components of the engine during a simulation is displayed in Fig.~\ref{traject}. It is observed in Fig.~\ref{traject}(a) that the piston bounces against the stops on each side of the engine while it collides with the particle. In the crossing events, in which the particle goes through the pore, one observes that the trajectory of the particle crosses that of the piston without bouncing back. Similarly, Fig.~\ref{traject}(b) shows how the piston pushes the hooks against the gravitational field by means of elastic collisions until the hook reaches the opposite end of the engine and then it is reset.\\  

\begin{figure}[!ht]
\centering
\includegraphics[scale=0.055]{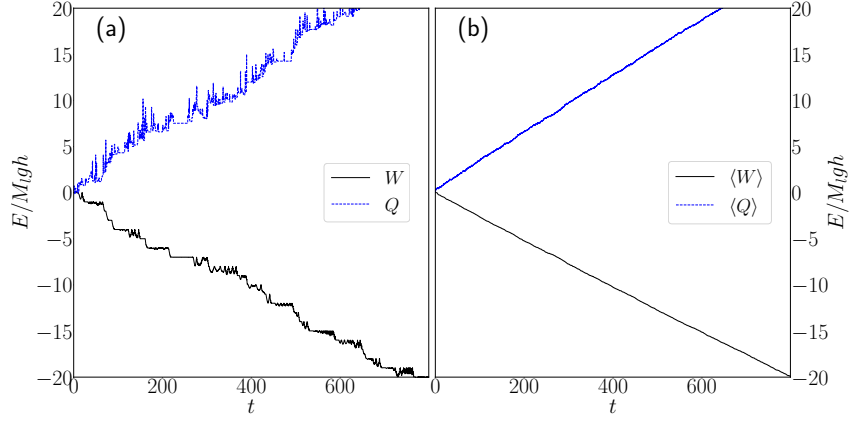}
\caption{Work and heat vs. time. Work done to lift the load mass ($W$) and heat exchanged between the particle and the thermal walls ($Q$) during a simulation of the engine where $p=0.2$ and $M_l=1.0$. (a) Single simulation. (b) Ensemble average.}
\label{qwtime}
\end{figure}

Figure~\ref{qwtime}(a) shows the accumulated work done on the load mass and the heat exchanged between the thermal walls and the particle as a function of time for a single simulation. Note that in this simple model the amount of work done per cycle by lifting the load mass is constant and equal to $M_lgh$, where $h=0.7$ corresponds to the height that the mass is lifted restricted by the chosen geometry of the container, which can be seen in the plot as discrete jumps in the accumulated work. That the work per cycle is constant means that for fixed $M_l$ the power delivered in a cycle ($M_lgh/t_c$) is a function of $t_c$ only. The oscillations of the work observed over the base line correspond to the bounces of the hooks observed in Fig.~\ref{traject}(b). Figure~\ref{qwtime}(b) shows the ensemble average of the heat and the work.\\

\begin{figure}[!ht]
 \centering
 \includegraphics[scale=0.055]{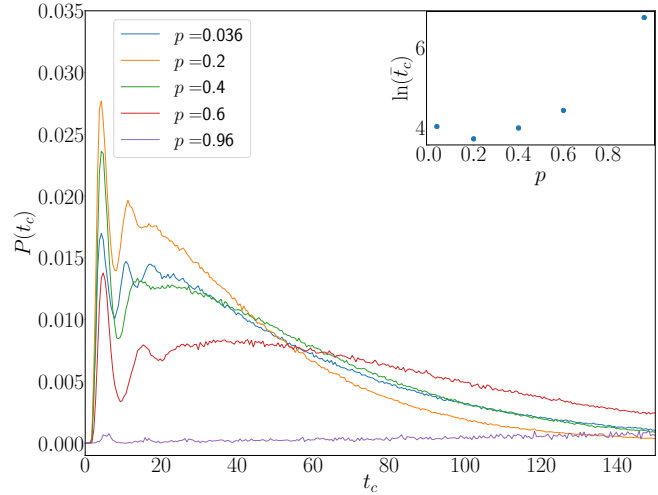}
 \caption{Cycle duration $t_c$ pdf for systems with $M_l=1$ and several values of $p$. The inset shows the logarithm of average cycle duration time $\bar{t}_c$ as a function of $p$.}
 \label{tvspor}
\end{figure}

\begin{figure}[!hb]
 \centering
 \includegraphics[scale=0.055]{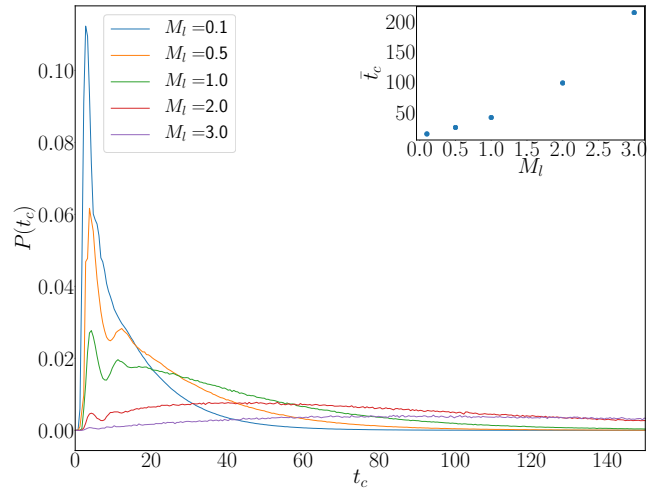}
 \caption{Cycle duration $t_c$ pdf for systems with $p=0.2$ and several values of $M_l$. The inset shows the average cycle duration time $\bar{t}_c$ as a function of $M_l$.}
 \label{tvsMass}
\end{figure}

\begin{figure}[!h]
 \centering
 \includegraphics[scale=0.055]{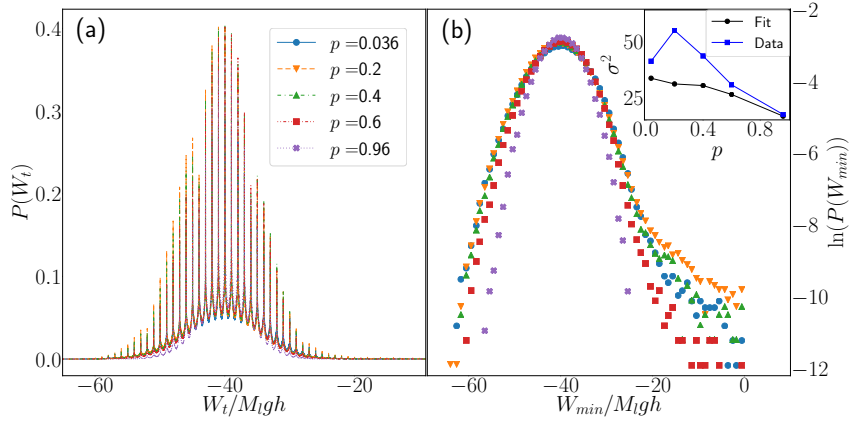}
 \caption{Work distribution at fixed load mass. (a) Pdf of the work performed by the engine on the load mass during a time window $t=40\bar{t}_c$ for systems with $M_l=1$ and several values of $p$. (b) Logarithm of the local minima of the curves shown in panel (a). Inset: Variance of the distribution computed from a Gaussian fit to the data around its maximum (squares) and from the numerical computation using all the work data (circles).}
 \label{Wvspor}
\end{figure}

\begin{figure}[!h]
 \centering
 \includegraphics[scale=0.055]{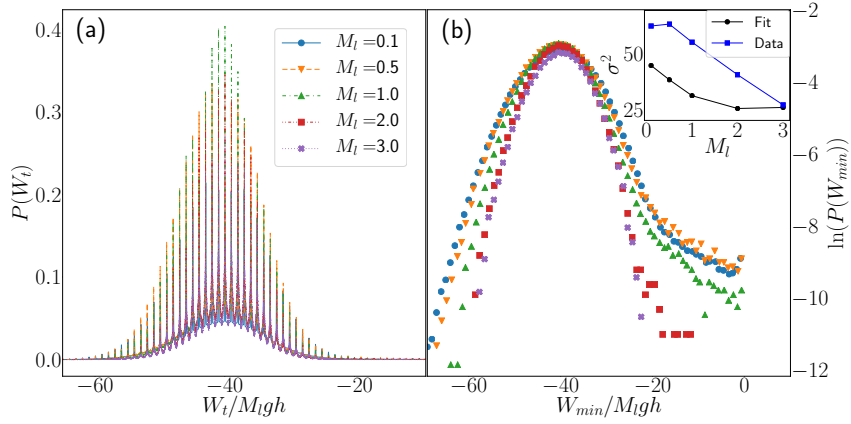}
 \caption{Work distribution at fixed porosity. (a) Pdf of the work performed by the engine on the load mass during a time window $t=40\bar{t}_c$ for systems with $p=0.2$ and several values of $M_l$. (b) Logarithm of the local minima of the curves shown in panel (a). Inset: Variance of the distribution computed from a Gaussian fit to the data around its maximum (squares) and from the numerical computation using all the work data (circles).}
 \label{WvsMass}
\end{figure}

The probability density function (pdf) of the time per cycle ($t_c$) for a fixed load mass $M_l=1$ and various porosities is displayed in Fig.~\ref{tvspor}. The pdf presents several peaks before decaying. In contrast with systems like, for example, a gas within a piston that follows an isothermal process, where it is compressed from state $A$ to state $B$ and then expanded from $B$ to $A$ to complete a cycle taking always the same time, the maxima found in the pdf can be explained by the fact that many different events might occur within one of the cycles of the engine with porous piston. For example, the piston can simply couple a mass soon after it was introduced and pull it all the way up, or it might wiggle for a time, bounce against it without coupling to it and pick instead the mass in the opposite side of the container. These different types of cycle might have different characteristic times that reflect in the pdf as different peaks. The inset of Fig~\ref{tvspor} shows that the average time ($\bar{t}_c$) gets to a minimum value near $p=0.2$ and as the work delivered per cycle is constant in this setup, it can be said that the engine operates at maximum power for a value close to $p=0.2$. \\

Similarly, Fig.~\ref{tvsMass} shows the pdf of $t_c$ for a fixed porosity $p=0.2$ and various load mass values. In contrast with its dependence on $p$, the average cycle duration as a function of $M_l$ increases monotonically. \\

On the other hand, we obtained the pdf of the work performed by the engine on the load mass during a time window of length $t$. Figure~\ref{Wvspor}(a) shows the pdf of the work $W_t$ normalized by $M_lgh$ when a time window of length $t=40\bar{t}_c$ is used for a fixed load mass and several values of the porosity. The periodic peaks observed are located at integer multiples of the work done during a single cycle. That the mode of the distributions in Fig.~\ref{Wvspor}(a) is close to $W_t/M_lgh \approx -40$ tells us simply that the most probable scenario when the engine runs during a time that spans 40 times the average cycle duration, is for it to complete 40 cycles.\\

In microscopic models of the  Carnot engine it has been found that the pdf of the work presents a long tail towards zero, but it approximates a Gaussian in the quasistatic limit\cite{Proesmans2015}, which sometimes has been used to approximate the distribution of the work\cite{Polettini2015,Martinez2016}. To estimate the importance of these tails in the case of the engine with porous piston, we obtained a curve consisting of all the local minima observed in Fig.~\ref{Wvspor}(a) and computed its logarithm, as shown in Fig.~\ref{Wvspor}(b). It is observed that the distribution deviates from a Gaussian as it presents the mentioned tail towards zero.\\ 

The inset in Fig.~\ref{Wvspor}(b) presents the variance of the distributions obtained directly from the data, as well as from a Gaussian fit performed around the mode of the distribution. The variance values computed from the data are considerably higher that those obtained from the fit because of the large tail towards zero of the distribution, but this difference becomes smaller as $p\rightarrow 0$. Also, it is observed that the variance of the data gets to a maximum around the point at which the engine operates at maximum power.\\

Figure~\ref{WvsMass}(a) shows the pdf of $W_t$, when a time window of length $t=40t_c$ was used, for a fixed porosity and several values of the load mass. It is observed that, when normalized by $M_lgh$, the amount of work done depends slightly on the load mass. As $M_l$ increases, the variance of the distribution decreases and it can be better fitted by a Gaussian function. As the pdf of $t_c$ is skewed towards large times and the accumulated work ($W_t$) increases as more cycles fit within the time window, the pdf of $W_t$ is skewed towards small absolute values. Figure~\ref{WvsMass}(b) shows the logarithm of the pdf of $W_t$, where a deviation from a Gaussian behavior is also observed.\\

A closely related quantity to the variance of the work distribution is the entropy production. It has been shown for various systems that the latter is bounded from below by the inverse of the so-called precision, i.e., the ratio of the variance to the squared mean value of a current,\cite{Barato2015,Pietzonka2017,Horrowitz2020}
\begin{align}
    \frac{\Sigma_t}{k_B}\geq\frac{2\langle W_t\rangle^2}{\text{Var}(W_t)},
    \label{tur}
\end{align}
where $\Sigma_t$ indicates the average entropy production during a time window of length $t$ and the current measured is the work done by the engine on the load masses. Table~\ref{preci} shows the precision as a function of averaging time $t/\bar{t}_c$. It is observed that for larger time windows there is less uncertainty in the value of the work. Assuming that the uncertainty relation given by Eq.~(\ref{tur}) holds for this system, these precision values indicate a lower bound for the entropy production.\\

\begin{table}[h]
    \centering
    \begin{tabular}{cc}
    \hline
        $t/\bar{t}_c$ & $2\left<W_t\right>^2/\mbox{Var}(W_t)$ \\
        \hline
        0.1 & 0.22 \\
        0.2 & 0.46 \\
        0.5 & 0.79 \\
        1.0 & 2.59 \\
        2.0 & 5.13 \\
        3.0 & 7.41 \\
        5.0 & 11.76 \\
        \hline
    \end{tabular}
    \caption{Inverse of the precision as a function of the averaging time $t/\bar{t}_c$ for a system with $M_l=1$ and $p=0.2$.}
    \label{preci}
\end{table}

\begin{figure}[!h]
 \centering
 \includegraphics[scale=0.055]{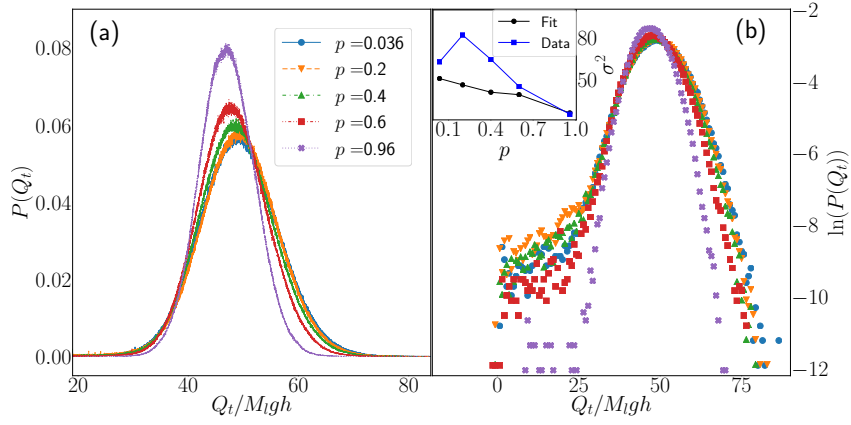}
 \caption{Heat distribution at fixed load mass. (a) Pdf of the heat exchanged between the particle and the thermal walls during a time window $t=40\bar{t}_c$ for systems with $M_l=1$ and several values of $p$. (b) Logarithm of the data of shown in panel (a). Inset: Variance of the distribution computed from a Gaussian fit to the data around its maximum (squares) and from the numerical computation using all the heat data (circles).}
 \label{Qvspor}
\end{figure}

\begin{figure}[!h]
 \centering
 \includegraphics[scale=0.055]{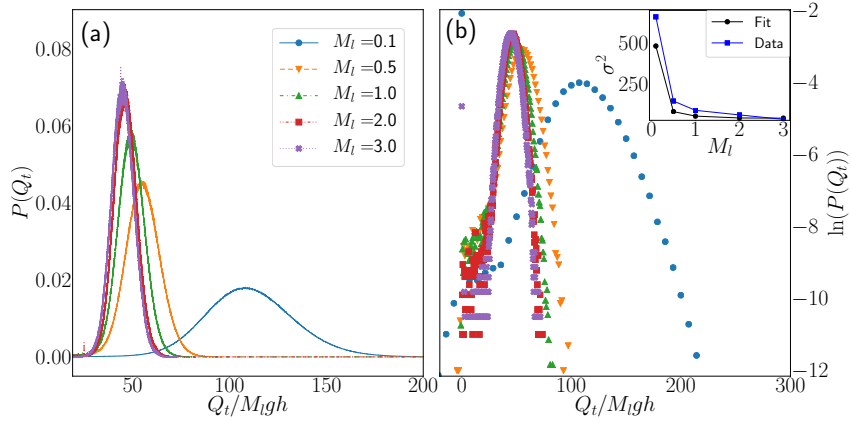}
 \caption{Heat distribution at fixed porosity. (a) Pdf of the heat exchanged between the particle and the thermal walls during a time window $t=40\bar{t}_c$ for systems with $p=0.2$ and several values of $M_l$. (b) Logarithm of the data shown in panel (a). Inset: Variance of the distribution computed from a Gaussian fit to the data around its maximum (squares) and from the numerical computation using all the heat data (circles).}
 \label{QvsMass}
\end{figure}

Figures~\ref{Qvspor} and \ref{QvsMass} show the pdf of the heat exchanged between the particle and the thermal walls.  As the change in the accumulated heat is not done in discrete steps as those present on the accumulated work, the peaks observed at multiples of $M_lgh$ in the work distribution are absent and, more importantly, the mode of the distributions for the heat varies noticeably as the parameters $p$ and $M_l$ are changed. It is this variation in heat absorption which produces the variation in the efficiency, as presented in the following paragraphs.\\

\begin{figure}[!h]
 \centering
 \includegraphics[scale=0.055]{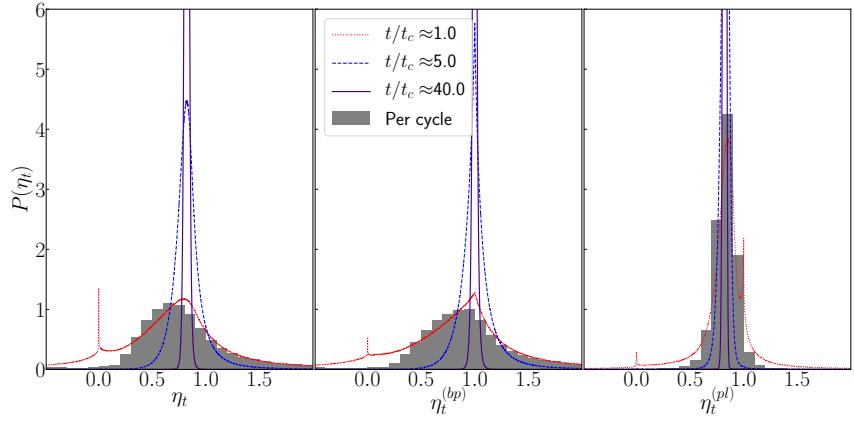}
 \caption{Pdf of the efficiency obtained from an ensemble of systems with $p=0.2$ and $M_l=1$. The bar plots refer to the actual efficiency per cycle of the engine, while the lines represent the efficiency pdf obtained by sampling several time windows of size $t$.}
 \label{effipdf}
\end{figure}

Figure~\ref{effipdf} shows an example of the pdf of $\eta_t$, $\eta_t^{(bp)}$ and $\eta_t^{(pl)}$ obtained from the simulations. The bar plot refers to the actual efficiency per cycle of the engine, while the lines show the efficiency pdf computed for several time windows of size $t$. The distributions observed are heavily skewed towards small values of $t$, while both the spread and skewedness decrease as $t\rightarrow\infty$. The peak at $\eta=0$ shows that for values of $t$ near $\bar{t}_c$ there is a high probability of the window falling within the same cycle doing almost no work, as shown in FIG.~\ref{qwtime}(a). Note that the shape of these distributions is different from those of previously studied systems,\cite{Gingrich2014,Verley2014,Proesmans2015B} in that in most of those cases two maxima are present. The presence of the two maxima can be explained by the possibility for the systems to function either as an engine or a heat pump with certain probability\cite{Polettini2015}. However, due to the reset of the masses, the engine with porous piston cannot operate in a reverse fashion. The sudden change in the probability distribution of microstates that the reset produces  amounts to a symmetry breaking,\cite{Roldan2014} and it has been observed that such events might affect the distributions and fluctuation relations that the system obeys.\cite{Buffoni2022}\\

\begin{figure}[!h]
 \centering
 \includegraphics[scale=0.055]{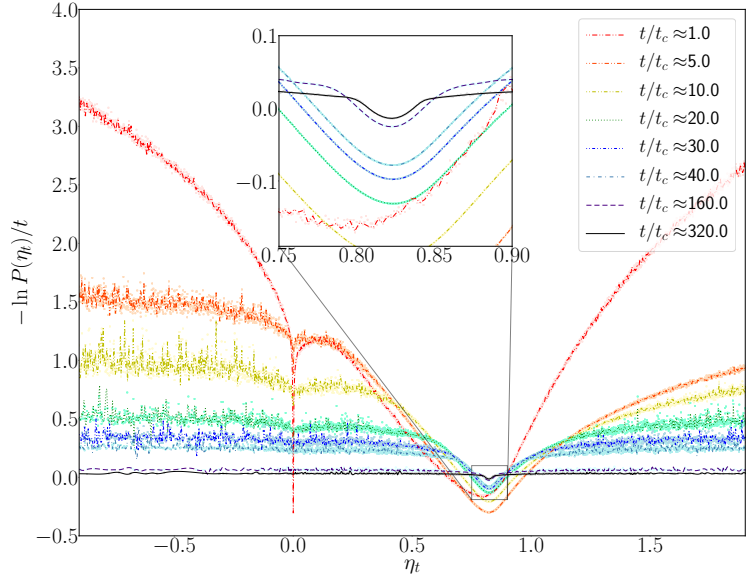}
 \caption{Estimation of the large deviation function. Behavior of the function $-\ln(P(\eta_t))/t$ in Eq. (\ref{ldf}) computed at $p=0.2$ and $M_l=1.0$, using several time window sizes.}
 \label{finiteldf}
\end{figure}

As explained in Section \ref{simulations}, the macroscopic efficiency $\bar{\eta}$ was computed by means of the large deviation function (Eq.~(\ref{ldf})) using several finite values of $t$ and then extrapolating the asymptotic behavior. Figure~\ref{finiteldf} shows Eq.~(\ref{ldf}) as $t$ goes from $\bar{t}_c$ to $320\bar{t}_c$ for an ensemble of systems with parameters $p=0.2$ and $M_l=1.0$. The values for $\bar{\eta}$ are shown  in Fig.~\ref{finiteldf} and were computed as an average of extrapolations using all possible combinations of ordered triplets of the curves,  as described in Ref.~\citenum{Proesmans2015B}.\\

\begin{figure}[!h]
 \centering
 \includegraphics[scale=0.055]{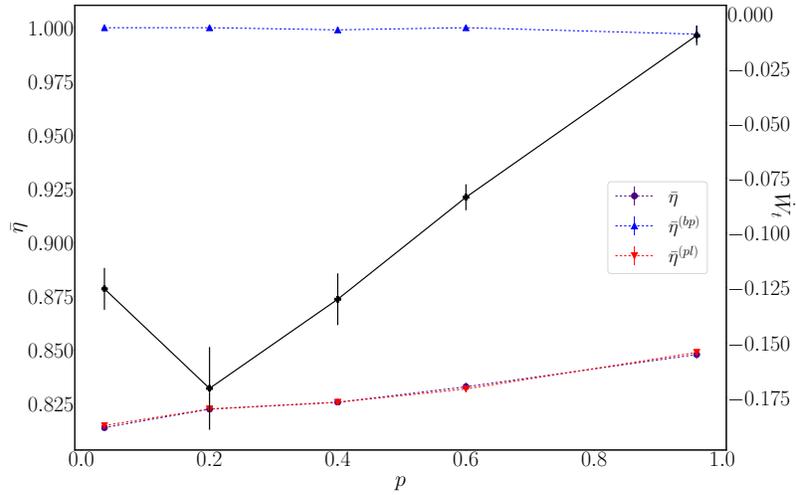}
 \caption{Macroscopic efficiency as a function of $p$ for ensembles of systems with $M_l=1$. The solid line plots the average power as a function of $p$ using a time window of $t=320\bar{t}_c$ (note the corresponding scale on the right side of the figure).}
 \label{effvsp}
\end{figure}

\begin{figure}[!h]
 \centering
 \includegraphics[scale=0.055]{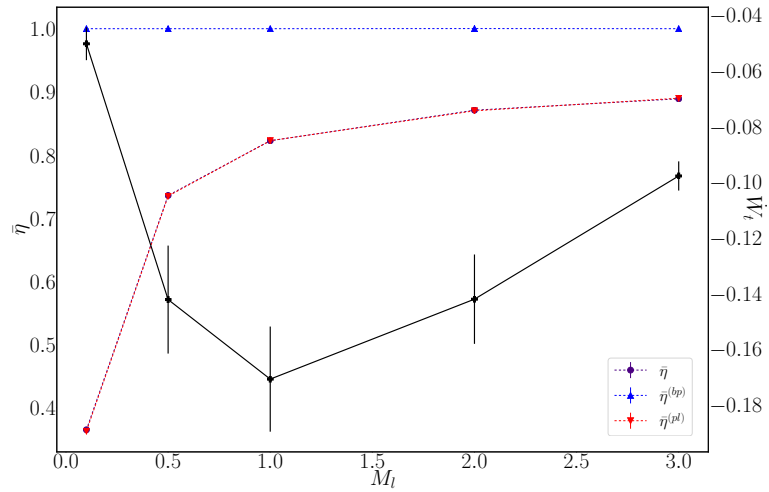}
 \caption{Macroscopic efficiency as a function of $M_l$ for ensembles of systems with $p=0.2$. The solid line plots the average power as a function of $M_l$ using a time window of $t=320\bar{t}_c$ (note the corresponding scale on the right side of the figure).}
 \label{effvsload}
\end{figure}

Figures~\ref{effvsp}~and~\ref{effvsload} plot the macroscopic efficiencies $\bar{\eta}$, $\bar{\eta}^{(bp)}$ and $\bar{\eta}^{(pl)}$ as functions of the porosity and load mass, respectively. It is observed that $\bar{\eta}^{(bp)}$ remains at 1, as on average all the energy that comes from the thermal walls goes to the piston. We also have that on average $\bar{\eta}=\bar{\eta}^{(bp)}\bar{\eta}^{(pl)}$, and that the dissipation is taking place as the work provided by the piston is transformed into potential energy of the load masses, because part of the energy transferred is converted into kinetic energy of the load masses that is lost at the instant of their removal from the system.\\
 
The associated average power is also shown in  Figs.~\ref{effvsp} and \ref{effvsload}, this time without the normalization by $M_lgh$. As it was previously observed, for fixed $M_l=1$ the engine operates at maximum power around $p=0.2$. On the other hand, for fixed $p=0.2$ the simulation data shows that the engine operates at maximum power for a load mass value around $M_l=1.0$.\\   

According to the general theory of feedback control in non-equilibrium systems,\cite{Sagawa2012} the work $W$ done on the system between two states satisfies the generalized Jarzynski relation\cite{Jarzynski1997, Sagawa2010} 
\begin{equation}
    \langle e^{-\beta(W-\Delta F)-I_c} \rangle = 1\,,
\end{equation}
where $\Delta F$ is the free energy difference between the initial and final state and $I_c$ is the mutual information between phase-space point $x$ and the outcome of its measure $y$. Here the measure is without error, therefore the average $\langle I_c \rangle$ has its maximum value equal to the Shannon information $H[X]$ of the trajectory $X=\{x(t)\}$
\begin{equation}
    \langle I_c \rangle = H[X] = - \int P[X] \ln P[X] \,dX  \,,
\end{equation}
with $P[X]$ is probability of the trajectory $X$ being realized. From the convexity of the exponential function one obtains 
\begin{equation}
    \langle W \rangle \geq \Delta F - k_B T\langle I_c \rangle
    \,.
\end{equation}
If in the initial and final state a load mass is hooked then $\Delta F=0$. In this point of view, the work that the machine is able to provide comes from the information obtained by measuring the piston position and performing the feedback control. Note that, contrarily to the Szilard engine, the position of the particle is not used in the feedback control.

\begin{figure}[!h]
 \centering
 \includegraphics[scale=0.055]{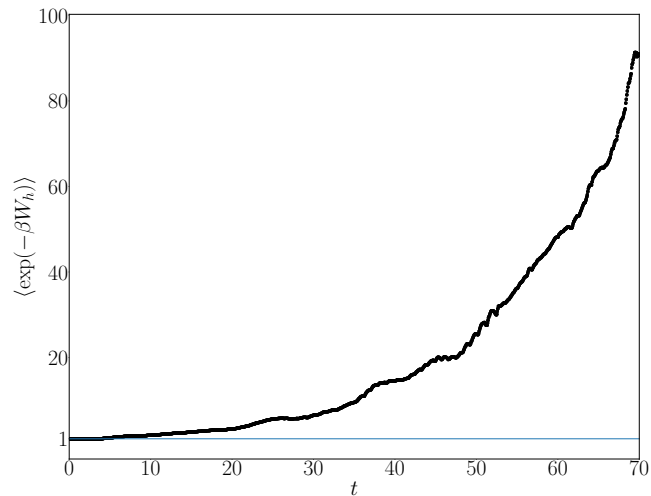}
 \caption{Estimation of the efficacy parameter ($\gamma$) as a function of time for a system with $p=0.2$ and $M_l=1$.}
 \label{efficacy}
\end{figure}

To characterize how efficient the feedback control is, one can introduce the efficacy parameter\cite{Sagawa2012} $\gamma$, which we obtain from our simulations and the relation
\begin{equation}
     \langle e^{-\beta(W-\Delta F)} \rangle = \gamma\,.
\end{equation}
For systems without feedback control $\gamma=1$, whereas for systems with feedback control $\gamma>1$.
Figure \ref{efficacy} shows the efficacy, estimated as the average of $\exp(-\beta W_h)$ over the ensemble of simulations, with  $W_h$ the work done during the intervals where the control parameter $\lambda$ is either $R$ or $L$, so that $\Delta F=0$. As expected from a feedback controlled system, the efficacy starts at 1 and increases from there as the time increases.

\section{Discussion}

By modifying Szilard's engine with the replacement of the piston by a porous one, the engine can convert heat into work without the need to store information about the position of the particle with respect to the piston. If the engine is not coupled to an external system on which to perform work, one can expect it to wiggle constantly as it exchanges heat back and forth with the thermal walls. However, the coupling of the engine with another system on which to do work introduces the possibility for energy dissipation.\\

We chose to couple the engine to a system similar to the one proposed in Szilard's original model, where work is done by lifting a small mass against a gravitational field. Energy is dissipated because once the load mass is removed from the system one does not take away merely the potential energy gained during the process, but also any kinetic energy that the load mass happens to have at that instant. 
This reset operation amounts to a symmetry breaking,\cite{Roldan2014,Buffoni2022} and as with engines in which an autonomous agent ultimately has to discard heat to a low temperature reservoir,\cite{Freitas2021} in this case the mechanical agent that places the new, low-energy load masses with zero kinetic energy, acts as a low temperature reservoir at $T_{\rm low}=0$ in the sense that it takes away all the kinetic energy from them. 
In addition, we chose to use a feedback control by unhooking and hooking a new load mass when the piston passes through positions $a$ and $b$ respectively. As observed previously in a thermal ratchet system,\cite{Sagawa2012} such a feedback allows for an efficacy parameter $\gamma>1$. In this case, the feedback reduces the amount of time that the system of spends with no mass hanging from the piston, enabling a greater power output.\\

If it were possible to take the system to a quasi-static limit at which the masses arrive to the extraction point with a velocity close to zero, one could expect that the amount of dissipated energy would also be very small. Although the system is not driven by an external parameter which can be made to change infinitely slowly, it is observed that as the load mass is increased, keeping all the other parameters constant, the efficiency of the engine does increase while the power decreases past a certain point. The effect of the porosity of the piston on the efficiency seems to be more modest, achieving slightly better efficiency and less power output at higher values of $p$. Finally, an approximate maximum in the power output of the engine was observed as a function of both, the load mass and the porosity of the piston. The power decreases as one moves away from these maxima in both directions, towards small and large values of parameters $p$ and $M_l$.


\section*{Data availability}
The code and datasets used and/or analysed during the current study are available from the corresponding author on reasonable request. 

\bibliography{sample}

%

\section*{Acknowledgements}

G. T. acknowledges support from Fondo de Investigaciones de la Facultad de Ciencias  de la Universidad de los Andes INV-2021-128-2267. M.C. thanks Vicerrector\'ia de Ciencia, Tecnolog\'ia e Innovación (VCTI) -- Universidad Antonio Nari\~no for financial support through Project 2022023. C.E.A. acknowledges financial support from Universidad del Rosario. The present article has been published in Scientific Reports: \href{https://www.nature.com/articles/s41598-022-18057-3}{Sci. Rep. {\bf12}, 13896 (2022)}.
 
%

\section*{Author contributions statement}

C.E.A. and M.C. performed the simulations. All authors analyzed the results and reviewed the manuscript. 

%
%
%
%
%

\end{document}